\newif\ifpublic
\publictrue 

\ifpublic
\documentclass[10pt,a4paper]{article}
\else
\documentclass[10pt,draft,a4paper]{article}
\fi

\usepackage[utf8]{inputenc} 
\usepackage[T1]{fontenc}    
\usepackage{electrum}
\usepackage{amsmath}       
\usepackage{amssymb}
\usepackage{amsthm}
\usepackage{nicefrac}       
\usepackage{fullpage}
\usepackage{mathpazo}
\usepackage{mathtools}
\usepackage{tikz}
\usetikzlibrary{calc}
\usetikzlibrary{decorations.pathreplacing, calligraphy}
\usepackage{thmtools}
\usepackage{thm-restate}
\usepackage{xcolor}
\usepackage{float}
\usepackage{caption}
\ifpublic
\usepackage[disable]{todonotes}
\else
\usepackage[colorinlistoftodos]{todonotes}
\fi

\usepackage[pagebackref,colorlinks=true,linkcolor=blue,urlcolor=blue,citecolor=blue,pdfstartview=FitH]{hyperref}
\usepackage[capitalize,nameinlink]{cleveref}

\usepackage{dsfont}
\usepackage{xstring}
\usepackage{xargs}

\newcommandx{\mc}[2]{%
  \IfStrEq{#1}{}%
    {\IfStrEq{#2}{}
        {M_\mathcal{C}}
        {M_\mathcal{C}(.,#2)}
    }
    {\IfStrEq{#2}{}
        {M_\mathcal{C}(#1,.)}
        {M_\mathcal{C}(#1,#2)}
    }
}

\newcommand{\rc}{R_{\mathcal{C}}}
\newcommand{\qklhashcode}{$q$-perfect hash code}
\newcommand{\rbdtrifferentcodes}{$r$-bounded trifferent code}
\newcommand{\densityrbd}{$r$-bounded density}
\newcommand{\boundeddeficit}{$r$-bounded deficit}
\newcommand{\limitingdeficit}{$\sup$-bounded deficit}

\newcommand{\F}{\mathbb{F}}

\newcommand{\calC}{\mathcal{C}}

\newcommand{\define}[1]{\textsb{#1}}

\newcommand{\set}[1]{\{#1\}}

\renewcommand{\Vec}[1]{\mathbf{#1}}

\newcommand{\bcap}{\mathsf{cap}}

\newtheorem{theorem}{Theorem}[section]

\newtheorem{corollary}{Corollary}[theorem]

\newtheorem{lemma}[theorem]{Lemma}

\newtheorem{definition}{Definition}[section]

\newtheorem{remark}[theorem]{Remark}

\DeclareMathOperator*{\E}{\mathbb{E}}
\DeclareMathOperator{\1}{\mathbf{1}}

\title{Improved Upper Bound for the Size of a Trifferent Code}

\date{}
\author{Siddharth Bhandari \thanks{Siddharth Bhandari is at the Toyota Technological Institute at Chicago. Email: \texttt{siddharth@ttic.edu}}, Abhishek Khetan \thanks{Abhishek Khetan is a post-doctoral researcher at the Tata Institute of Fundamental Research, CAM, Bangalore. Email: \texttt{khetan21@tifrbng.res.in}} }

\begin{document}

\maketitle

\begin{abstract}
    A subset $\calC\subseteq\set{0,1,2}^n$ is said to be a \emph{trifferent} code (of block length $n$) if for every three distinct codewords $x,y, z \in \calC$, there is a coordinate $i\in \set{1,2,\ldots,n}$ where they all differ, that is, $\set{x(i),y(i),z(i)}$ is same as $\set{0,1,2}$.
    Let $T(n)$ denote the size of the largest trifferent code of block length $n$.
    Understanding the asymptotic behavior of $T(n)$ is closely related to determining the zero-error capacity of the $(3/2)$-channel defined by Elias~\cite{Elias1988}, and is a long-standing open problem in the area.
    Elias had shown that $T(n)\leq 2\times (3/2)^n$ and prior to our work the best upper bound was $T(n)\leq 0.6937 \times (3/2)^n$ due to Kurz~\cite{Kurz2023trifferent}.
    We improve this bound to $T(n)\leq c \times n^{-2/5}\times (3/2)^n$ where $c$ is an absolute constant.
\end{abstract}



\section{Introduction}
\label{section:introduction}

Let $q$ be a positive integer and let  $\Sigma = \set{0, 1, 2, \ldots, q-1}$ be a finite alphabet.
We use the notation $[n]$ to denote the set $\set{1, 2, \ldots, n}$ when $n$ is a positive integer.

\begin{definition}[$q$-perfect hash codes \& Trifferent codes]
	\label{defn:qhashcode}
	For positive integers $q\geq 2$ and $n$, a code $\mathcal{C} \subseteq \Sigma^n$ is said to be a $q$-\define{perfect hash code} of \define{block length} $n$ if for any $q$ distinct codewords $x_{1}, x_{2}, \ldots, x_{q}$ in $\mathcal{C}$ we have a coordinate $i\in [n]$ such that $\set{x_{j}(i)\mid 1\leq j\leq q} = \Sigma$, where $x(i)$ denotes the $i^{th}$ coordinate of $x$.
	When $q=3$, a \qklhashcode{} is also referred to as a \define{trifferent} code.
 
	We will write $T(q,n)$ to denote the maximum size a \qklhashcode{} of block length $n$ attains.
\end{definition}

Understanding the asymptotics of $T(q,n)$ as $n$ increases is an important question, both in information theory and computer science. 
In this paper we will focus solely on the case of $q=3$.
As mentioned in \cref{defn:qhashcode}, in this case a \qklhashcode{}  is popularly referred to as a `trifferent' code and examining the growth of $T(3,n)$ is referred to as the 'trifference problem', a long-standing open problem that has garnered considerable attention.
(See for instance the $2014$ Shannon Lecture: \href{https://www.itsoc.org/video/isit-2014-mathematics-distinguishable-difference}{`On The Mathematics of Distinguishable Difference'} by János Körner.)
Since we concern ourselves only with the case of $q=3$ in the remainder of the paper, we refer to $T(3,n)$ as $T(n)$. In a seminal work Elias~\cite{Elias1988} showed that $T(n)\leq {2\times (3/2)^n}$.
Prior to our work the best upper bound on $T(n)$ was $T(n)\leq 0.6937 \times (3/2)^n$ for $n\geq 10$, due to Kurz~\cite{Kurz2023trifferent}. We improve this bound for sufficiently large $n$ in the following result.
\begin{restatable}[Main theorem]{theorem}{maintheorem}
	\label{thm:main_thm}
	There exists a universal constant $c$ with the following property.
	Let  $\calC\subseteq \set{0,1,2}^n$ be a trifferent code of block length $n$ as defined in \cref{defn:qhashcode}.
	Then, $|\calC| \leq c \times n^{-2/5}\times (3/2)^n$.
	Thus, $T(n)\leq c \times n^{-2/5}\times (3/2)^n$.
\end{restatable}

Before delving into the trifference problem, we elucidate how \qklhashcode{s} are connected to perfect hashing and how they simultaneously serve as error-correcting codes for a classical channel in information theory. Subsequently, we highlight notable findings in the estimation of $T(q,n)$ for the cases when $q>3$.

\subsection*{General $q$-perfect Hash Codes}
A \qklhashcode{} $\calC = \set{x_{1},x_{2},\ldots,x_{s}}$ of block length $n$ and size $s$ is readily seen to be a family $\set{h_1, \ldots, h_n}$ of $n$ hash functions from  $[s]$ to $\Sigma$ where the $i^{th}$ hash function $h_i$ maps $j\in [s]$ to $x_j(i)$, i.e., the $i^{th}$ coordinate of the codeword $x_j$.
The family of hash functions $\set{h_i}$ has the property that any subset $Q$ of size $q$ of the domain $[s]$  is perfectly hashed by at least one of the hash functions $h_i$, i.e., $\set{h_i(j)\colon j\in Q} = \Sigma$.
Alternatively, a \qklhashcode{}, say $\calC$ as described above, can also be cast as a cover of the $q$-uniform complete hypergraph on the vertex set $[s]$, say $K_s(q)$,  using $n$ hypergraphs which are $q$-uniform and $q$-partite.
Specifically, we think of each hash function $h_i$ as a hypergraph $H_i$ whose vertex set is $[s]$ and the edge set is $\set{Q\subseteq [s] : |Q|=q \land \set{h_i(j) \mid j\in Q} = \Sigma}$. See the excellent survey of Radhakrishnan~\cite{Radhakrishnan2001EntropyAC} for more details.

A \qklhashcode{} also serves as an error-correcting code for a classical channel studied in zero-error information theory: the $q/(q-1)$ channel.
The input and output alphabets of this channel are a set of $q$ symbols, namely $\Sigma$; when the channel receives the symbol $i \in \Sigma$ as input, the output symbol can be anything other than $i$ itself.
For the $q/(q-1)$ channel it is impossible to recover the message without error if the code has at least two codewords: in fact, no matter how large the block length, for every set of up to $q-1$ input codewords, one can construct an output word that is compatible with \emph{all} of them.
However, there exist codes with positive rate where on receiving an output word from the channel, one can narrow down the possibilities for the input message to a set of size at most $q-1$, that is, we can \emph{list-decode} with lists of size $q-1$.
Such codes are called $(q-1)$-list-decoding codes for the $q/(q-1)$ channel.
It is well known that a \qklhashcode{} $\calC$ of block length $n$ and size $s$ is equivalent to a $(q-1)$-list-decoding code for the $q/(q-1)$ channel with block length $n$ (for instance see the introduction of Bhandari and Radhakrishnan \cite{BhandariR2022_IEEE_TOIT}).

\begin{definition}[Rate \& Capacity]
	\label{defn:gen_rate_capacity}
	For positive integers $q\geq 2$ and $n$, let $\calC$ be a \qklhashcode{} of block length $n$.
	Following Elias~\cite{Elias1988}, we define the \define{rate} of $\calC$ as $\rc\coloneqq \frac{1}{n} \log_2 (|\calC|/(q-1))$.
	We define the $q$-\define{capacity} as 
\[ \bcap (q) = \limsup_{n \rightarrow \infty}  \frac{1}{n} \log_2 \frac{T(q,n)}{q-1}.\]
\end{definition}

\begin{remark}
It is not known if `$\limsup$' can be replaced by `$\lim$' in the definition of capacity; see \cite[Footnote~1]{Arikan1994}.
\end{remark}



Many significant improvements have been made in understanding $\bcap(q)$ and related quantities for $q > 3$.
We list some of them below and refer the reader to the work of Bhandari and Radhakrishnan \cite{BhandariR2022_IEEE_TOIT} for a more detailed survey. 
Fredman and Koml\'{o}s' seminal work \cite{FredmanK1984-perfect} established $\bcap(q) \leq \exp(-B_1q)$ for a constant $B_1>0$, independent of $q$. 
Guruswami and Riazanov \cite{GuruswamiR2019} demonstrated the non-optimality of the Fredman-Komlós upper bound for $q\geq 4$ and provided explicit improvements for $q=5,6$. Costa and Dalai \cite{CostaD2020} resolved a conjecture by Guruswami and Riazanov, completing the explicit computation for improving the Fredman-Koml\'{o}s bound across all $q$, and introduced an alternative method yielding substantial enhancements for $q=5,6$. 
For $q=4$, Dalai, Guruswami, and Radhakrishnan \cite{DalaiGR2020} improved the upper bound to $\bcap(4) \leq 6/19 \approx 0.3158$, surpassing Arikan's previous bound of $0.3512$ \cite{Arikan1994}, while K\"{o}rner and Marton \cite[eq (1.2)]{KornerM1988} established a lower bound of $\bcap(4) \geq (1/3)\lg (32/29)\approx 0.0473$. 
Additionally, Xing and Yuan \cite{XingY2019} extended K\"{o}rner and Marton's concatenation technique, demonstrating improved lower bounds on capacity for $q = 4,8$, all odd integers greater than $3$ and less than $25$, and sufficiently large $q$ not congruent to $2\pmod{4}$.

\subsection*{The Trifference Problem}
Despite receiving considerable attention, progress for the trifference problem has been relatively modest when compared to the situation for $q>3$.
Elias~\cite{Elias1988} showed that $0.08 \approx\lg (3) - 1.5 \leq \bcap(3) \leq \lg(3) -1 \approx 0.58$; 
{K\"{o}rner and Marton~\cite{KornerM1988} improved the lower bound above to $0.212 \approx (1/4)\lg (9/5)\leq \bcap(3)$ via code concatenation.}
Under the further assumption of linearity, i.e., if we think of $\Sigma$ as $\F_3$ and assume that the trifferent code $\calC\subseteq \F_3^n$ is a linear subspace of $\F_3^n$, some improvements have been obtained in the upper bound on $\bcap(3)$. Pohoata and Zakharov~\cite{PohoataZ22} obtained ${\mathsf{linear\text{-}cap}}(3)\leq (1/4-\epsilon)\times \log_3(2) \approx 0.3962 - \epsilon$ for some absolute constant $\epsilon>0$ following which Bishnoi, D'haeseleer, Gijswijt and Potukuchi~\cite{BishnoiDGP23} obtained ${\mathsf{linear\text{-}cap}}(3)\leq  (1/4.55)\times \log_3(2)\approx 0.3483$.

Notwithstanding the above results, the current best upper bound on $\bcap(3)$ for general trifferent codes remains the one given by Elias~\cite{Elias1988}, up to a constant factor. 
As such, there has been an impetus to view $T(n)$, the largest size of a trifferent code of block length $n$, with a more refined lens.
Elias' upper bound and K\"{o}rner and Marton's lower bound can be recast in terms of $T(n)$ as $T(n)\leq 2\times (3/2)^n$ and $T(n)\geq (9/5)^{n/4}$ respectively.
Recently, via a computer search for a large trifferent code of block length up to $n \leq 6$, combined with a number theoretic argument, Fiore, Gnutti and Polak~\cite{FIOREGP2022} showed that $T(n) \leq 1.09\times (3/2)^n$ for $n\geq 12$.
Even more recently Kurz~\cite{Kurz2023trifferent} extended the computer search for trifferent codes of block lengths up to $n\leq 7$ and obtained $T(n)\leq 0.6937\times (3/2)^n$ for $n\geq 10$. 

What makes studying $T(n)$ intriguing is the fact that the upper bound of $2\times (3/2)^n$ is obtained via a relatively simple pruning argument (described below) which has proved difficult to improve.
(See for instance the work of Costa and Dalai~\cite{COSTAD2021} as to the limits of the `slice rank' method for the trifference problem, which, however was successful in bounding the largest size of a $3$-AP free set in $\F_3^n$).
Additionally, the lower bound of $(9/5)^{n/4}$ is obtained not via a purely random construction, but via concatenating a random outer code and an algebraic inner code (known as the Tetra code).
The pruning argument for the upper bound of $2\times (3/2)^n$ is as follows: let $\calC$ be a trifferent code of block length $n$ and let $a_1\in \set{0, 1, 2}$ be a least occurring symbol in the first coordinate of all the codewords.
Then, let $\calC_1$ be the code obtained by deleting those codewords from $\calC$ that have $a_1$ in the first coordinate.
Observe that $|\calC_1| \geq (2/3)|\calC|$.
Now since $\mathcal C$ was trifferent, same is true for $\mathcal C_1$.
But any three distinct strings from $\mathcal C_1$ cannot exhibit the trifference property in the first coordinate and hence for any three distinct codewords $x, y, z$ in $\calC_1$ there must exist a coordinate $i > 1$ such that $\set{x(i),y(i),z(i)} = \set{0,1,2}$.
Proceeding iteratively in this manner we let $a_2$ be a least occurring symbol in the second coordinate of codewords in $\calC_1$, and then obtain $\calC_2$ from $\calC_1$ by deleting those codewords which have $a_2$ in their second coordinate, and so on, till we obtain $\calC_n$.
Thus, $|\calC_n|\geq (2/3)^n\times|\calC|$.
But observe that $\mathcal C_n$ is a trifferent code where three distinct strings cannot exhibit the trifference property in \emph{any} coordinate.
Therefore $2\geq |\mathcal C_n|$, which leads to $|C|\leq 2\times (3/2)^n$.

We restate our main result below, which is an improved upper bound on $T(n)$.
\maintheorem*

To prove \cref{thm:main_thm} we first introduce a close variant of trifferent codes which we call `bounded trifferent' codes. 

\begin{definition}[\rbdtrifferentcodes{s}]
\label{defn:rbdtriffcodes}
	Let $\calC\subseteq\set{0,1,2}^n$ be a trifferent code of block length $n$.
	For an integer $r\geq 0$, we call $\calC$ an $r$-\define{bounded trifferent code} if for all codewords $x\in \calC$ we have that the number of $2$'s in $x$ is $r$, i.e., $|\set{i\in [n]\colon x(i)=2}|= r$. 
	Further, for $n\geq r$ let $T_b(n,r)$ denote the maximum size an \rbdtrifferentcodes{} of block length $n$ attains.
\end{definition}

In the remainder of the paper, when we talk about $T_b(n,r)$ it is to be understood that $n\geq r$ as an \rbdtrifferentcodes{} of block length $n< r$ has size $0$.
Note that $T_b(n,r)\leq 2\times \binom{n}{r} $ as for a given subset of coordinates $S\subseteq[n]$, in any trifferent code of block length $n$ there can be at most two codewords, say $x$ and $y$, such that $S$ is precisely the location of $2$'s for both $x$ and $y$, i.e., $S = \set{i\in [n]\mid x(i) = 2} = \set{i\in [n]\mid y(i) = 2}$.
If there were three, they would be a counter-example to the trifference property.
Next, we prove a simple lemma relating $T(n)$ and $T_b(n,r)$, thus, highlighting the importance of studying \rbdtrifferentcodes{s}.

\begin{lemma}[Size of trifferent codes in terms of \rbdtrifferentcodes{s}]
\label{lem:size_relation_triff_rbdtriff}
	$$T(n)\leq 2^{(r-n)} \times \frac{T_b(n,r)}{\displaystyle \binom{n}{r}} \times 3^n$$
\end{lemma}
\begin{proof}
	It will be convenient to think of $\Sigma := \set{0, 1, 2}$ as $\F_3$.
	Let $\calC$ be a trifferent code of block length $n$.
	Let $A_r$ denote the set of all those $x\in \Sigma^n$ such that $x$ witnesses exactly $r$ $2$s.
	More precisely
	\begin{equation*}
		A_r = \set{x\in \F_3^n:\ |\set{i\in [n]:\ x(i) = 2}| = r}.
	\end{equation*}
    For strings $x$ and $v$ in $\F_3^n$ let $x+v$ denote addition in $\F_3^n$.
    Observe that the code $\calC+v \coloneqq \set{x+v \mid x\in \calC }$ is also trifferent for any string $v\in \F_3^n$. Hence,   $|(\calC+v)\cap A_r|\leq T_b(n, r)$. 

	Let $\mathbf v$ be a uniformly random element of $\F_3^n$.
	We write $\calC+\Vec{v}$ to denote the random subset $\calC + {v}$ where $v$ is picked according to the distribution of $\Vec{v}$.
	Note that for any $x\in \F_3^n$ we have:
	\begin{equation*}
		\E_{\mathbf v}[\1[x\in \calC + \Vec{v}]]
            = \mathbb P_{\mathbf v}(x - \mathbf v \in \mathcal C)
            =\frac{|\calC|}{3^n}.
	\end{equation*}
	Therefore,
	\begin{equation*}
		T_b(n, r)
		\geq \E_{\mathbf v}[|(\calC + \Vec{v})\cap A_r|]
		= \sum_{x\in A_r} \E_{\mathbf v} [\1[x \in \calC + \Vec{v} ]] = |A_r|\times\frac{|\mathcal C|}{3^n}.
	\end{equation*}
	Now using $|A_r| = \binom{n}{r} 2^{n-r}$, we get the desired result.
\end{proof}

\begin{remark}
    Even more generally, the above lemma shows that for any $S\subseteq\set{0,1,2}^n$ if $T_b(S)$ denotes that largest size of a trifferent code contained in $S$, then we have $T(n)\leq \frac{T_b(S)}{|S|} \times 3^n$.
	Hence, it might also be interesting to look at sets $S$ which are not of the form $\set{x\in \set{0,1,2}^n \colon \#2's \text{ in } x= r}$ for some integer $r$.
\end{remark}

In light of the above lemma it is natural to define the notion of an $r$-bounded density.

\begin{definition}[\densityrbd{}]
\label{defn:density_rbounded}
	Recall that $T_b(n,r)$ denotes the maximum size an \rbdtrifferentcodes{} of block length $n$ attains.
	The $r$-\define{bounded density} at block length $n$, denoted as $\rho_b(n,r)$, is defined as $$\rho_b(n, r) = 2^{(r-n)} \times \frac{T_b(n,r)}{\binom{n}{r}}$$
\end{definition}

Hence, \cref{lem:size_relation_triff_rbdtriff} can be cast as $T(n)\leq \rho_b(n,r) \times 3^n$ for all $r\geq 0$.
The bound obtained using the pruning argument, that is $T(n)\leq 2\times (3/2)^n$, can be obtained as an instantiation of \cref{lem:size_relation_triff_rbdtriff} with $r=0$.
In this case, it is clear that $T_b(n,r)=2$ (there can be at most two codewords in a trifferent code if the symbol $2$ does not appear in any codeword) and hence $\rho_b(n,r)=2^{1-n}$. It turns out that $\rho_b(n,1) = 2^{2-n}$ as we show that $T_b(n,1)=2n$ (see \cref{lem:one_bounded_triff_codes}).
This bound is worse than what we obtained on $\rho_b(n,0)$. However, the situation improves when we consider $r\geq 2$. 
Our main contribution is showing that for $r = 3$, $\rho_b(n,r)\leq c\times n^{-2/5}\times 2^{-n}$ for some absolute constant $c$.
More precisely, we have the following result.

\begin{restatable}[Bounding \densityrbd{}]{theorem}{denstiyrbounded}
	\label{thm:bounding_denstiy_rbounded}
	Let $T_b(n,r)$ and $\rho_b(n,r)$ be as defined in \cref{defn:rbdtriffcodes,defn:density_rbounded} respectively. Then, we have constants $c'$ and $c$ such that  
	\begin{enumerate}
		\item[a)] $T_b(n,2)\leq c'\times n^{5/3}$ and hence $\rho_b(n,2)\leq c\times n^{-1/3}\times 2^{-n}$ and
		\item[b)] $T_b(n,3)\leq c'\times n^{13/5}$  and hence $\rho_b(n,3)\leq c\times n^{-2/5}\times 2^{-n}$.
	\end{enumerate}
\end{restatable}

We describe the proof idea of \cref{thm:bounding_denstiy_rbounded} for the case when $r=2$.
We proceed via constructing a graph related to an \rbdtrifferentcodes{}, say $\calC_b$, with block length $n$.
This graph has roughly as many edges as $|\calC_b|$.
The crucial observation is that certain bipartite structures are forbidden in this graph. Then, an application of the famous K\H{o}v\'{a}ri--S\'{o}s--Tur\'{a}n (KST) theorem yields a bound on the number of edges in this graph which also serves as a bound on the size of $\calC_b$.
We give the details below.

Recall that each codeword in $\calC_b$ has exactly two $2$'s.
Now, consider the graph $G_{\calC_b}$ on the vertex set $[n]$ where for each codeword $x\in \calC_b$ an edge $\set{i,j}$ with $i\neq j$ is added to $G_{\calC_b}$ if $x(i)=x(j)=2$, i.e., $i$ and $j$ are the locations of $2$'s in $x$. 
Note that an edge $\set{i,j}$ can be added by at most $2$ codewords in $\calC_b$.
Hence, $G_{\calC_b}$ has at least half as many edges as $|\calC_b|$.
Next, we show via the PHP and trifference property of $\calC_b$ that $K_{3,9}$---the complete bipartite graph with the partite sets having sizes $3$ and $9$ respectively---is forbidden in $G_{\calC_b}$.
Applying the KST theorem yields an upper bound on the number of edges in $G_{\calC_b}$ as $c''\times n^{5/3}$ for some constant $c''$.
Hence, we obtain our desired bound on $|\calC_b|$ and $T_b(n,2)$.

The proof for when $r=3$ proceeds along similar lines but we define the graph $G_{\calC_b}$ more prudently.
The detailed proof of \cref{thm:bounding_denstiy_rbounded} appears in \cref{sec:upper_bd_rbdtrifferentcodes}.

Armed with \cref{thm:bounding_denstiy_rbounded,lem:size_relation_triff_rbdtriff} we easily obtain the proof of \cref{thm:main_thm}.

\begin{proof}[Proof of \cref{thm:main_thm}]
	With $r=3$ we have the following.
	\begin{align*}
		T(n)&\leq 2^{(r-n)} \times \frac{T_b(n,r)}{\binom{n}{r}} \times 3^n \qquad \text{by \cref{lem:size_relation_triff_rbdtriff}}\\
		& = \rho_b(n,r) \times 3^n \\
		& \leq c\times n^{-2/5}\times 2^{-n} \qquad \text{ by \cref{thm:bounding_denstiy_rbounded}.}
	\end{align*}
    giving the desired result.
\end{proof}

From the above discussion it is tempting to analyze $\rho_b(n,r)$ when both $r$ and $n$ are growing with $n$ growing much faster than $r$. As such, we define a notion of \boundeddeficit{} and \limitingdeficit{} which serve to get a sense of the speed at which $\rho_b(n,r)$ decays.

\begin{definition}[\boundeddeficit{} \& ${\sup}$-{bounded deficit}]
	For an integer $r\geq 1$, let $T_b(n,r)$ denote the maximum size an \rbdtrifferentcodes{} of block length $n$ attains. Let $\Delta_r(n) = r-\frac{\log{T_b(n,r)}}{\log{n}}$. Then, the ${r}$-\define{bounded deficit} is defined as
	\begin{align*}
	    \Delta_r&\coloneqq\limsup_{n\to \infty}\left(r-\frac{\log{T_b(n,r)}}{\log{n}}\right)\\
        &= \limsup_{n\to \infty} \Delta_r(n).
	\end{align*}
	and the ${\sup}$-\define{bounded deficit} is defined as
	$$
		\Delta_\infty\coloneqq\lim_{r\to \infty}\Delta_r.
	$$ 
\end{definition}

\begin{remark}
	\begin{enumerate}
		\item To be more precise we should have defined $\Delta_\infty$ as $\limsup_{r\to \infty}\Delta_r$.
			However, \cref{cor:bouding_densityrbounded} shows that $\Delta_r$ is increasing in $r$.
		\item  If we unpack the above definition in terms of $T_b(n,r)$, then \cref{lem:size_relation_triff_rbdtriff} yields that $$T(n)\leq c_r\times n^{-\Delta_r(n)}\times (3/2)^n$$ where $c_r$ is a constant depending only on $r$. Hence, studying $\Delta_r$ as $r$ grows is helpful in proving better upper bounds on $T(n)$. 
	\end{enumerate}
\end{remark}

Since, $T_b(n,r)$ is at most $2\times \binom{n}{r}$, the \boundeddeficit{}, $\Delta_r$, is always non-negative.
In fact, by \cref{thm:bounding_denstiy_rbounded} we have shown that $\Delta_2\geq 1/3$ and $\Delta_3\geq 2/5$. 
Via a simple application of the PHP we show that $\Delta_r$ is increasing in $r$.
\begin{corollary}
	\label{cor:bouding_densityrbounded}
    $\Delta_{r+1}\geq \Delta_{r}$ for any integer $r\geq 1$.
	Hence, for all integers $r\geq 3$, there are constants $c_r$ and $c_r'$ such that we have: $T_b(n,r)\leq c_r'\times n^{r-2/5}$ and hence $\rho_b(n,r)\leq c_r\times n^{2/5}\times 2^{-n}$.
\end{corollary}
\begin{proof}
    Let $\calC_b\subseteq \set{0,1,2}^n$ be an $r+1$-bounded trifferent code of block length $n$.
    By double-counting we can find a coordinate $i\in [n]$ such that there is a subset $\calC_b'\subseteq\calC_b$ of size at least $r/n \times |\calC_b|$ and every codeword of $\calC_b'$ has a $2$ at coordinate $i$. Notice that $\calC_b'$ can be equivalently thought of as an \rbdtrifferentcodes{} with block length $n-1$ since the $i$-th coordinate is same (namely 2) for each codeword. Hence, $T_b(n,r+1)\leq n/r \times T_b(n-1,r)$ from where it follows that $\Delta_{r+1}\geq \Delta_{r}$.
\end{proof}

Next, we turn our attention to establish upper bounds on $\Delta_r$, i.e., prove lower bounds on $T_b(n,r)$. Our constructions are based on thinking about the codewords as point-line incidences in an appropriate finite-dimensional vector space over a finite field. See \cref{sec:lower_bounds_rbdtrifferent_codes} for the detailed constructions.

\begin{restatable}[Upper bounds on $\Delta_r$]{theorem}{upperboundondeltar}
\label{thm:upper_bound_on_deltar}
    $T_b(n,1)= 2n$ and hence $\Delta_1 =0$. 
    Also, $\Delta_3 \leq 3/2$, i.e., $T_b(n,3)\geq c_1\times n^{3/2}$ for some constant $c_1>0$: further, for every positive integer $r$, a power of $3$, we have $\Delta_r\leq r-r^{\alpha}$ where $\alpha = 1- \log_{3}(2) \approx 0.369$.
\end{restatable}

\subsection*{Further Questions}

A few interesting questions which arise are as follows. Is $\Delta_\infty = \infty$? If yes, this immediately shows that for any constant $d$ we have $T(n)\leq n^{-d}\times (3/2)^n$, i.e., the size of a family of trifferent codes of growing block lengths, say $n \to \infty$, decays faster than $(3/2)^n$ divided by any polynomial factor.
A more nuanced understanding in terms of $\rho_b(n,r)$ with growing $r$ might also lead to an improved upper bound on $\bcap(3)$, hence making progress on the long-standing open problem.
Further, it is also interesting to understand $\Delta_r$ more precisely for small values of $r$ such as $2$ and $3$: is $\Delta_2=1/2$?

\section*{Acknowledgements}
    We are highly indebted to Prof. Jaikumar Radhakrishnan for various insightful discussions and for painstakingly proofreading the manuscript. 
We thank Prof. Alexander Razborov for helpful discussions. 
We also acknowledge Aakash Bhowmick for helping us with experiments related to constructing trifferent codes. 
Lastly, we thank Prof. Nishant Chandgotia for his inputs and encouragement.

\section{Upper bounds on the size of \rbdtrifferentcodes{s}}
\label{sec:upper_bd_rbdtrifferentcodes}

In this section we prove \cref{thm:bounding_denstiy_rbounded}.
We restate the theorem below for convenience. 

\denstiyrbounded*

To prove \cref{thm:bounding_denstiy_rbounded} we will need to apply the famous result of K\H{o}v\'{a}ri, S\'{o}s and Tur\'{a}n, known popularly as the KST theorem, or rather a version of it due to Hyltén-Cavallius.

\begin{theorem}[KST theorem due to Hyltén-Cavallius~\cite{HyltenCavallius1958}]
\label{thm:KST}
The Zarankiewicz function $z(u,v;s,t)$ denotes the maximum possible number of edges in a bipartite graph $G=(U\cup V, E)$ for which $|U|=u$ and $|V|=v$, but which does not contain a subgraph of the form $K_{s,t}$ where $s$ vertices come from $U$ and $t$ from $V$ (here $K_{s,t}$ denotes the complete bipartite graph with $s$ and $t$ vertices in the two partite sets).
Then, $$z(u,v;s,t) < (t-1)^{\frac{1}{s}} (u-s+1) v^{1-\frac{1}{s}} + (s-1)v.$$

\end{theorem}

\begin{proof}[Proof of \cref{thm:bounding_denstiy_rbounded}]
	We first proceed with the proof when $r=2$. Let $\calC_b$ be an \rbdtrifferentcodes{} of block length $n$ with $r=2$.
	Thus, every codeword of $\calC_b$ has two $2$'s.
	As discussed previously, we construct a graph $G_{\calC_b}$ using the code $\calC_b$.
	The graph $G_{\calC_b}$ has the vertex set $[n]$ and the set of edges $E$ is
	$$
		E = \set{\set{i,j}\mid i\neq j \land \exists x\in \calC_b: x(i)=x(j)=2}.
	$$
	In other words, for each codeword $x\in \calC_b$, an edge, say $x_e = \set{i,j}$, is added to $G_{\calC_b}$, where $i$ and $j$ are the locations of $2$'s in $x$.

	As we have argued previously, for any subset $S\subseteq [n]$ of coordinates there can be at most two codewords, say $x$ and $y$, such that $S$ is precisely the location of $2$'s for both $x$ and $y$, i.e., $S=\set{i\in [n]\mid x(i)=2}=\set{i\in [n]\mid y(i)=2}$. (If there were three, they would contradict the trifference property.) 
	This shows that an edge $\set{i,j}$ can be added by at most $2$ codewords in $\calC_b$.
	Hence,
	$$
		|E|\geq |\calC_b|/2,
	$$
	i.e., there are at least half as many edges as $|\calC_b|$.
	At the cost of another factor of $1/2$ on $|E|$ we can assume that $G_{\calC_b}$ is bipartite of the form $(U\cup V, E')$, where $V = [n]\setminus U$ and $|E'| \geq |E|/2$.
	(A random equi-bipartition of $[n]$ will have this property on expectation.)

	Next, we show that $K_{3,9}$ is forbidden in $G_{\calC_b}$.
	To see this suppose there exist distinct $i_1,i_2,i_3$ in $U$ and distinct $j_1,j_2,\ldots,j_{9}$ in $V$ such that subgraph induced by $G_{\calC_b}$ on these vertices is $K_{3,9}$.
	We denote the edge $\set{i_k,j_\ell}$ by $e_{k,\ell}$.
	Further, let $x_{k,\ell}$ denote a codeword in $\calC_b$ corresponding to which the edge $e_{k,\ell}$ was added to $G_{\calC_b}$: if there are two such codewords then choose one arbitrarily and fix it.
	By the PHP there is a subset $T\subseteq[9]$ of size at least $3$ such that all codewords in $\set{x_{1,\ell}\mid \ell \in T}$ have the same value on the coordinates $i_2$ and $i_3$, i.e.,  $|\set{x_{1,\ell}(i_2)\mid \ell \in T}|=1$ and  $|\set{x_{1,\ell}(i_3)\mid \ell \in T}|=1$.
	Again, by the PHP we can find a subset $T'\subseteq T$ of size at least $2$ such that all codewords in $\set{x_{2,\ell}\mid \ell \in T'}$ have the same value on the coordinate $i_3$, i.e.,  $|\set{x_{2,\ell}(i_3)\mid \ell \in T'}|=1$.
    WLOG let us assume that $T' = \set{j_1, j_2}$ (see Figure \cref{fig:illustration for r 2}).

    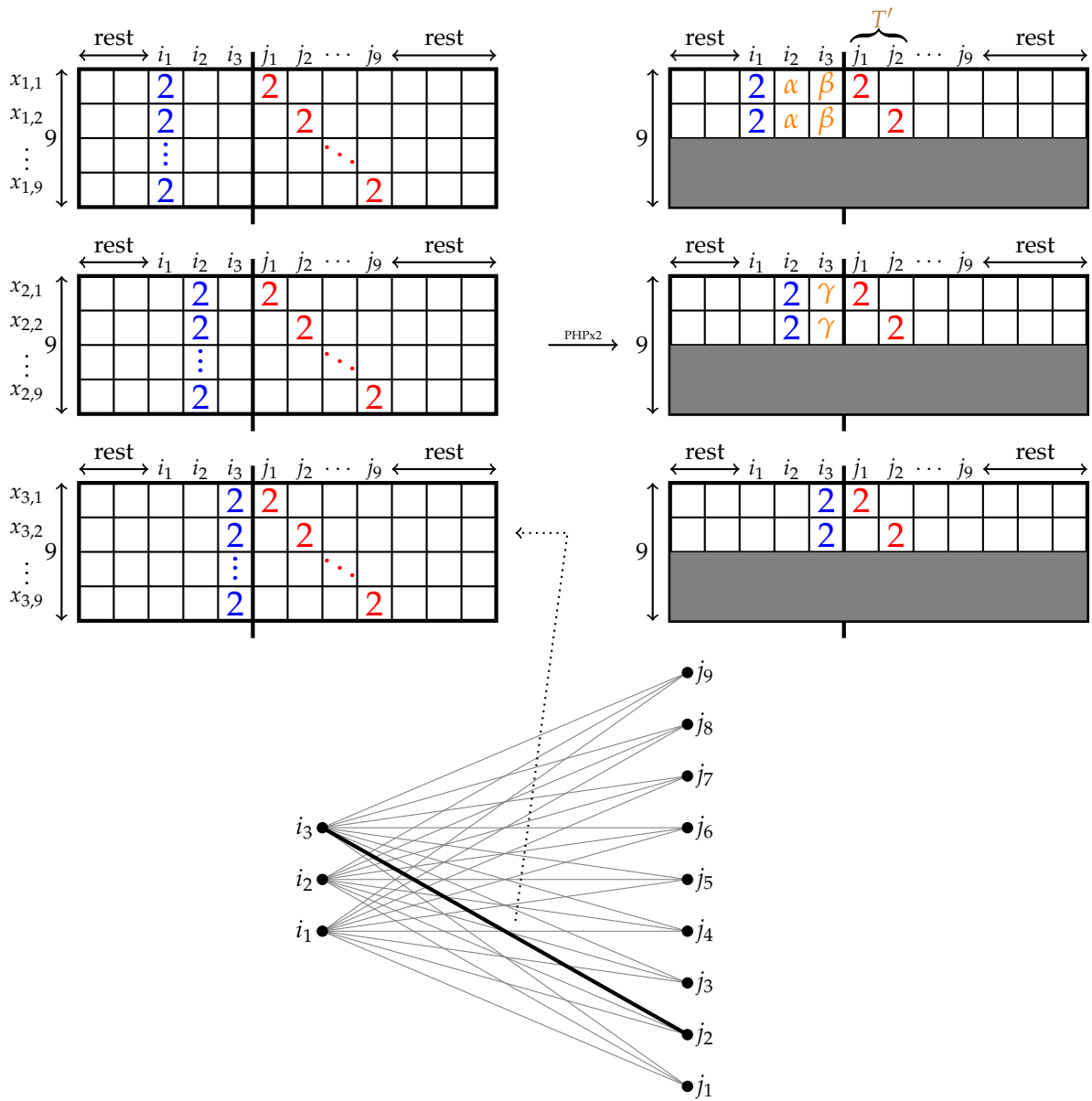
\begin{figure}
		\centering
		\def\babyoffset{0.4}
		\def\offset{0.5}
		\def\columns{4}

		\begin{tikzpicture}[scale=0.5]
			\begin{scope}[xshift=-1cm, yshift=0]
				\begin{scope}
					\draw[ultra thick] (0,0) rectangle (12,4);

					\draw[thick] (5,0) -- (5,4);

					\foreach \row in {1,...,3}
					\draw[thick] (0,4-\row) -- (12,4-\row);

					\foreach \col in {1,...,11}
					\draw[thick] (\col,0) -- (\col,4);

					\draw[ultra thick] (5, 0 - \offset) -- (5, 4 + \offset);

					\draw[<->, thick] (-\offset, 0) -- (-\offset, 4);
					\node at (-0.8, 2) {9};

					\draw[<->, thick] (0, 4 + \babyoffset) -- (2, 4 + \babyoffset);
					\node at (1, 4 + \offset + \babyoffset) {rest};

					\draw[<->, thick] (9, 4 + \babyoffset) -- (12, 4 + \babyoffset);
					\node at (10.5, 4 + \offset + \babyoffset) {rest};

					\node[font=\small] at (2 + \offset, \columns + \babyoffset) {$i_1$};
					\node[font=\small] at (3 + \offset, \columns + \babyoffset) {$i_2$};
					\node[font=\small] at (4 + \offset, \columns + \babyoffset) {$i_3$};

					\node[font=\small] at (5 + \offset, \columns + \babyoffset) {$j_1$};
					\node[font=\small] at (6 + \offset, \columns + \babyoffset) {$j_2$};
					\node[font=\small] at (7 + \offset, \columns + \babyoffset) {$\cdots$};
					\node[font=\small] at (8 + \offset, \columns + \babyoffset) {$j_9$};
				
					\def\rownum{2}

					\node[text=blue, font=\Large] at (\rownum + \offset, 0 + \offset) {$2$};
					\node[text=blue, font=\Large] at (\rownum + \offset, 1 + 1.5*\offset) {$\vdots$};
					\node[text=blue, font=\Large] at (\rownum + \offset, 2 + \offset) {$2$};
					\node[text=blue, font=\Large] at (\rownum + \offset, 3 + \offset) {$2$};

					\node[text=red, font=\Large] at (5 + \offset, 3 + \offset) {$2$};
					\node[text=red, font=\Large] at (6 + \offset, 2 + \offset) {$2$};
					\node[text=red, font=\Large] at (7 + \offset, 1 + 1.5*\offset) {$\ddots$};
					\node[text=red, font=\Large] at (8 + \offset, 0 + \offset) {$2$};

					\node[font=\small] at (0 - 3*\offset, 0 + 1.2*\offset) {$x_{1,9}$};
					\node at (0 - 3*\offset, 1 + 1.2*\offset) {$\vdots$};
					\node[font=\small] at (0 - 3*\offset, 2 + 1.2*\offset) {$x_{1, 2}$};
					\node[font=\small] at (0 - 3*\offset, 3 + 1.2*\offset) {$x_{1, 1}$};
				\end{scope}

				\begin{scope}[xshift=0, yshift=-6cm]
					\draw[ultra thick] (0,0) rectangle (12,4);

					\draw[thick] (5,0) -- (5,4);

					\foreach \row in {1,...,3}
					\draw[thick] (0,4-\row) -- (12,4-\row);

					\foreach \col in {1,...,11}
					\draw[thick] (\col,0) -- (\col,4);

					\draw[ultra thick] (5, 0 - \offset) -- (5, 4 + \offset);

					\draw[<->, thick] (-\offset, 0) -- (-\offset, 4);
					\node at (-0.8, 2) {9};

					\draw[<->, thick] (0, 4 + \babyoffset) -- (2, 4 + \babyoffset);
					\node at (1, 4 + \offset + \babyoffset) {rest};

					\draw[<->, thick] (9, 4 + \babyoffset) -- (12, 4 + \babyoffset);
					\node at (10.5, 4 + \offset + \babyoffset) {rest};

					\node[font=\small] at (2 + \offset, \columns + \babyoffset) {$i_1$};
					\node[font=\small] at (3 + \offset, \columns + \babyoffset) {$i_2$};
					\node[font=\small] at (4 + \offset, \columns + \babyoffset) {$i_3$};

					\node[font=\small] at (5 + \offset, \columns + \babyoffset) {$j_1$};
					\node[font=\small] at (6 + \offset, \columns + \babyoffset) {$j_2$};
					\node[font=\small] at (7 + \offset, \columns + \babyoffset) {$\cdots$};
					\node[font=\small] at (8 + \offset, \columns + \babyoffset) {$j_9$};

					\def\rownum{3}
					\node[text=blue, font=\Large] at (\rownum + \offset, 0 + \offset) {$2$};
					\node[text=blue, font=\Large] at (\rownum + \offset, 1 + 1.5*\offset) {$\vdots$};
					\node[text=blue, font=\Large] at (\rownum + \offset, 2 + \offset) {$2$};
					\node[text=blue, font=\Large] at (\rownum + \offset, 3 + \offset) {$2$};

					\node[text=red, font=\Large] at (5 + \offset, 3 + \offset) {$2$};
					\node[text=red, font=\Large] at (6 + \offset, 2 + \offset) {$2$};
					\node[text=red, font=\Large] at (7 + \offset, 1 + 1.5*\offset) {$\ddots$};
					\node[text=red, font=\Large] at (8 + \offset, 0 + \offset) {$2$};

					\node[font=\small] at (0 - 3*\offset, 0 + 1.2*\offset) {$x_{2, 9}$};
					\node at (0 - 3*\offset, 1 + 1.2*\offset) {$\vdots$};
					\node[font=\small] at (0 - 3*\offset, 2 + 1.2*\offset) {$x_{2, 2}$};
					\node[font=\small] at (0 - 3*\offset, 3 + 1.2*\offset) {$x_{2, 1}$};
				\end{scope}

				\begin{scope}[xshift=0, yshift=-12cm]
					\draw[ultra thick] (0,0) rectangle (12,4);

					\draw[thick] (5,0) -- (5,4);

					\foreach \row in {1,...,3}
					\draw[thick] (0,4-\row) -- (12,4-\row);

					\foreach \col in {1,...,11}
					\draw[thick] (\col,0) -- (\col,4);

					\draw[ultra thick] (5, 0 - \offset) -- (5, 4 + \offset);

					\draw[<->, thick] (-\offset, 0) -- (-\offset, 4);
					\node at (-0.8, 2) {9};

					\draw[<->, thick] (0, 4 + \babyoffset) -- (2, 4 + \babyoffset);
					\node at (1, 4 + \offset + \babyoffset) {rest};

					\draw[<->, thick] (9, 4 + \babyoffset) -- (12, 4 + \babyoffset);
					\node at (10.5, 4 + \offset + \babyoffset) {rest};

					\node[font=\small] at (2 + \offset, \columns + \babyoffset) {$i_1$};
					\node[font=\small] at (3 + \offset, \columns + \babyoffset) {$i_2$};
					\node[font=\small] at (4 + \offset, \columns + \babyoffset) {$i_3$};

					\node[font=\small] at (5 + \offset, \columns + \babyoffset) {$j_1$};
					\node[font=\small] at (6 + \offset, \columns + \babyoffset) {$j_2$};
					\node[font=\small] at (7 + \offset, \columns + \babyoffset) {$\cdots$};
					\node[font=\small] at (8 + \offset, \columns + \babyoffset) {$j_9$};

					\def\rownum{4}
					\node[text=blue, font=\Large] at (\rownum + \offset, 0 + \offset) {$2$};
					\node[text=blue, font=\Large] at (\rownum + \offset, 1 + 1.5*\offset) {$\vdots$};
					\node[text=blue, font=\Large] at (\rownum + \offset, 2 + \offset) {$2$};
					\node[text=blue, font=\Large] at (\rownum + \offset, 3 + \offset) {$2$};

					\node[text=red, font=\Large] at (5 + \offset, 3 + \offset) {$2$};
					\node[text=red, font=\Large] at (6 + \offset, 2 + \offset) {$2$};
					\node[text=red, font=\Large] at (7 + \offset, 1 + 1.5*\offset) {$\ddots$};
					\node[text=red, font=\Large] at (8 + \offset, 0 + \offset) {$2$};

					\node[font=\small] at (0 - 3*\offset, 0 + 1.2*\offset) {$x_{3, 9}$};
					\node at (0 - 3*\offset, 1 + 1.2*\offset) {$\vdots$};
					\node[font=\small] at (0 - 3*\offset, 2 + 1.2*\offset) {$x_{3, 2}$};
					\node[font=\small] at (0 - 3*\offset, 3 + 1.2*\offset) {$x_{3, 1}$};
				\end{scope}
			\end{scope}

			\begin{scope}[xshift=16cm, yshift=0]
				\draw[->, thick] (-3.5, -4) -- (-1.5, -4);
				\node[font=\tiny] at (-2.5, -3.7) {PHPx2};

				\begin{scope}
					\draw[ultra thick] (0,0) rectangle (12,4);

					\draw[thick] (5,0) -- (5,4);

					\foreach \row in {1,...,3}
					\draw[thick] (0,4-\row) -- (12,4-\row);

					\foreach \col in {1,...,11}
					\draw[thick] (\col,0) -- (\col,4);

					\draw[ultra thick] (5, 0 - \offset) -- (5, 4 + \offset);

					\draw[<->, thick] (-\offset, 0) -- (-\offset, 4);
					\node at (-0.8, 2) {9};

					\draw[<->, thick] (0, 4 + \babyoffset) -- (2, 4 + \babyoffset);
					\node at (1, 4 + \offset + \babyoffset) {rest};

					\draw[<->, thick] (9, 4 + \babyoffset) -- (12, 4 + \babyoffset);
					\node at (10.5, 4 + \offset + \babyoffset) {rest};

					\draw[decorate, decoration={calligraphic brace, amplitude=5pt}, ultra thick] (5 + 0.2, 5 - 0.2) -- (7 - 0.2, 5 - 0.2) node[midway,below=6pt] {$ $};
					\node[color=brown] at (6.1, 5.5) {$T'$};

					\node[font=\small] at (2 + \offset, \columns + \babyoffset) {$i_1$};
					\node[font=\small] at (3 + \offset, \columns + \babyoffset) {$i_2$};
					\node[font=\small] at (4 + \offset, \columns + \babyoffset) {$i_3$};

					\node[font=\small] at (5 + \offset, \columns + \babyoffset) {$j_1$};
					\node[font=\small] at (6 + \offset, \columns + \babyoffset) {$j_2$};
					\node[font=\small] at (7 + \offset, \columns + \babyoffset) {$\cdots$};
					\node[font=\small] at (8 + \offset, \columns + \babyoffset) {$j_9$};
				
					\def\rownum{2}

					\node[text=blue, font=\Large] at (\rownum + \offset, 0 + \offset) {$2$};
					\node[text=blue, font=\Large] at (\rownum + \offset, 1 + 1.5*\offset) {$\vdots$};
					\node[text=blue, font=\Large] at (\rownum + \offset, 2 + \offset) {$2$};
					\node[text=blue, font=\Large] at (\rownum + \offset, 3 + \offset) {$2$};

					\node[text=red, font=\Large] at (5 + \offset, 3 + \offset) {$2$};
					\node[text=red, font=\Large] at (6 + \offset, 2 + \offset) {$2$};
					\node[text=red, font=\Large] at (7 + \offset, 1 + 1.5*\offset) {$\ddots$};
					\node[text=red, font=\Large] at (8 + \offset, 0 + \offset) {$2$};

					\node[text=orange, font=\large] at (3 + \offset, 3 + \offset) {$\alpha$};
					\node[text=orange, font=\large] at (3 + \offset, 2 + \offset) {$\alpha$};

					\node[text=orange, font=\large] at (4 + \offset, 3 + \offset) {$\beta$};
					\node[text=orange, font=\large] at (4 + \offset, 2 + \offset) {$\beta$};

					\filldraw [fill=white!50!black, semitransparent] (0,0) rectangle (12,2);
				\end{scope}

				\begin{scope}[xshift=0, yshift=-6cm]
					\draw[ultra thick] (0,0) rectangle (12,4);

					\draw[thick] (5,0) -- (5,4);

					\foreach \row in {1,...,3}
					\draw[thick] (0,4-\row) -- (12,4-\row);

					\foreach \col in {1,...,11}
					\draw[thick] (\col,0) -- (\col,4);

					\draw[ultra thick] (5, 0 - \offset) -- (5, 4 + \offset);

					\draw[<->, thick] (-\offset, 0) -- (-\offset, 4);
					\node at (-0.8, 2) {9};

					\draw[<->, thick] (0, 4 + \babyoffset) -- (2, 4 + \babyoffset);
					\node at (1, 4 + \offset + \babyoffset) {rest};

					\draw[<->, thick] (9, 4 + \babyoffset) -- (12, 4 + \babyoffset);
					\node at (10.5, 4 + \offset + \babyoffset) {rest};

					\node[font=\small] at (2 + \offset, \columns + \babyoffset) {$i_1$};
					\node[font=\small] at (3 + \offset, \columns + \babyoffset) {$i_2$};
					\node[font=\small] at (4 + \offset, \columns + \babyoffset) {$i_3$};

					\node[font=\small] at (5 + \offset, \columns + \babyoffset) {$j_1$};
					\node[font=\small] at (6 + \offset, \columns + \babyoffset) {$j_2$};
					\node[font=\small] at (7 + \offset, \columns + \babyoffset) {$\cdots$};
					\node[font=\small] at (8 + \offset, \columns + \babyoffset) {$j_9$};

					\def\rownum{3}
					\node[text=blue, font=\Large] at (\rownum + \offset, 0 + \offset) {$2$};
					\node[text=blue, font=\Large] at (\rownum + \offset, 1 + 1.5*\offset) {$\vdots$};
					\node[text=blue, font=\Large] at (\rownum + \offset, 2 + \offset) {$2$};
					\node[text=blue, font=\Large] at (\rownum + \offset, 3 + \offset) {$2$};

					\node[text=red, font=\Large] at (5 + \offset, 3 + \offset) {$2$};
					\node[text=red, font=\Large] at (6 + \offset, 2 + \offset) {$2$};
					\node[text=red, font=\Large] at (7 + \offset, 1 + 1.5*\offset) {$\ddots$};
					\node[text=red, font=\Large] at (8 + \offset, 0 + \offset) {$2$};

					\node[text=orange, font=\large] at (4 + \offset, 3 + \offset) {$\gamma$};
					\node[text=orange, font=\large] at (4 + \offset, 2 + \offset) {$\gamma$};

					\filldraw [fill=white!50!black, semitransparent] (0,0) rectangle (12,2);
				\end{scope}

				\begin{scope}[xshift=0, yshift=-12cm]
					\draw[ultra thick] (0,0) rectangle (12,4);

					\draw[thick] (5,0) -- (5,4);

					\foreach \row in {1,...,3}
					\draw[thick] (0,4-\row) -- (12,4-\row);

					\foreach \col in {1,...,11}
					\draw[thick] (\col,0) -- (\col,4);

					\draw[ultra thick] (5, 0 - \offset) -- (5, 4 + \offset);

					\draw[<->, thick] (-\offset, 0) -- (-\offset, 4);
					\node at (-0.8, 2) {9};

					\draw[<->, thick] (0, 4 + \babyoffset) -- (2, 4 + \babyoffset);
					\node at (1, 4 + \offset + \babyoffset) {rest};

					\draw[<->, thick] (9, 4 + \babyoffset) -- (12, 4 + \babyoffset);
					\node at (10.5, 4 + \offset + \babyoffset) {rest};

					\node[font=\small] at (2 + \offset, \columns + \babyoffset) {$i_1$};
					\node[font=\small] at (3 + \offset, \columns + \babyoffset) {$i_2$};
					\node[font=\small] at (4 + \offset, \columns + \babyoffset) {$i_3$};

					\node[font=\small] at (5 + \offset, \columns + \babyoffset) {$j_1$};
					\node[font=\small] at (6 + \offset, \columns + \babyoffset) {$j_2$};
					\node[font=\small] at (7 + \offset, \columns + \babyoffset) {$\cdots$};
					\node[font=\small] at (8 + \offset, \columns + \babyoffset) {$j_9$};

					\def\rownum{4}
					\node[text=blue, font=\Large] at (\rownum + \offset, 0 + \offset) {$2$};
					\node[text=blue, font=\Large] at (\rownum + \offset, 1 + 1.5*\offset) {$\vdots$};
					\node[text=blue, font=\Large] at (\rownum + \offset, 2 + \offset) {$2$};
					\node[text=blue, font=\Large] at (\rownum + \offset, 3 + \offset) {$2$};

					\node[text=red, font=\Large] at (5 + \offset, 3 + \offset) {$2$};
					\node[text=red, font=\Large] at (6 + \offset, 2 + \offset) {$2$};
					\node[text=red, font=\Large] at (7 + \offset, 1 + 1.5*\offset) {$\ddots$};
					\node[text=red, font=\Large] at (8 + \offset, 0 + \offset) {$2$};

					\filldraw [fill=white!50!black, semitransparent] (0,0) rectangle (12,2);
				\end{scope}
			\end{scope}

			\begin{scope}[scale=1.5]
				\def\radius{0.1}
				\def\khiskaoright{4}
				\def\khiskaoupar{-15}

				\foreach \i in {1, 2, 3} {
					\coordinate (A\i) at (\khiskaoright, \i + \khiskaoupar);
				}

				\foreach \j in {1,2,3,4, 5, 6, 7, 8, 9} {
					\coordinate (B\j) at (7 + \khiskaoright, \j - 3 + \khiskaoupar);
				}

				\foreach \i in {1, 2, 3} {
					\foreach \j in {1, 2, 3, 4, 5, 6, 7, 8, 9} {
						\draw[color=gray] (A\i) -- (B\j);
					}
				}

				\foreach \i in {1, 2, 3}{
					\draw[fill=black] (A\i) circle (\radius);
					\node[left] at (A\i) {$i_\i$};
				}

				\foreach \j in {1, 2, 3, 4, 5, 6, 7, 8, 9}{
					\draw[fill=black] (B\j) circle (\radius);
					\node[right] at (B\j) {$j_\j$};
				}

				\draw[ultra thick] (A3) -- (B2);
				
				\coordinate (Mtemp) at ($(A3)!0.5!(B2)$);
				\coordinate (M) at ($(Mtemp) + (0.2, 0.2)$);

				\draw[->, dotted, thick] (M) -- ($(M) + (1, 7 + \offset)$) -- ($(M) + (0, 7 + \offset)$);
			\end{scope}
		\end{tikzpicture}

		\captionof{figure}{Illustrating the proof for $r = 2$. The collection of codewords before the application of the PHP is displayed in the left column of the figure: they correspond to the edges of the $K_{3,9}$ supposed to exist for the sake of contradiction. For example, the codeword $x_{3,2}$ corresponds to the highlighted edge $\set{i_3,j_2}$, and the first block of codewords on the left corresponds to the edges incident to $i_1$. After applying the PHP we are left with the non-grayed codewords in the right column corresponding to the set $T'=\set{j_1,j_2}$. Eventually, we find three codewords from the non-grayed ones which do not exhibit the trifference property. }
		\label{fig:illustration for r 2}
    \end{figure}
    
    Then, by the PHP there must exist two vertices $i_c$ and $i_d$ (with $c<d$) in $\set{i_1,i_2,i_3}$ such that $x_{c,1}(j_2)=x_{d,1}(j_2)$.
	But then the three codewords $x_{c,1},x_{d,1}$ and $x_{c,2}$ are a counter-example to the trifference of $\calC_b$.
	To see this let $w\in [n]$.
        If $w \notin \set{i_c,i_d,j_1, j_2}$, then none of the three codewords have the symbol $2$ at $w$ and hence $x_{c, 1}, x_{d, 1}$ and $x_{c, 2}$ cannot witness trifference at $w$.
        Now if $w=i_c$, then both $x_{c,1}$ and $x_{c,2}$ have the symbol $2$ at $w$; if $w=i_d$, then, because of our definition of $T'$ we have $x_{c,1}(w) = x_{c,2}(w)$; if $w=j_1$, then both $x_{c,1}$ and $x_{d,1}$ have the symbol $2$ at $w$; finally if $w=j_2$, then by our assumption $x_{c,1}(j_2)=x_{d,1}(j_2)$.
        So no matter what $w$ is, some two of $x_{c, 1}, x_{d, 1}$ and $x_{x, 2}$ agree on $w$, and hence these three codewords contradict the trifference property.
	Hence, $G_{\calC_b}$ is $K_{3,9}$-free.

	Applying \cref{thm:KST} with $u,v = n/2$ and $s=3,t=9$, yields $c''\times n^{5/3}$ as and upper bound on the size of $E'$ for some constant $c''$.
        Hence, we obtain our desired bound on $|\calC_b|$ and $T_b(n,2)$.

	Next, we turn our attention to the case when $r=3$.
	Again let $\calC_b$ be an \rbdtrifferentcodes{} with $r=3$ and block length $n$.
	Recall that each codeword in $\calC_b$ now has three $2$'s.
	This time we construct a bi-partite graph $G_{\calC_b} = (U,V,E)$ with $U=[n]$, $V = \binom{[n]}{2}$ and edge set $E$ described as follows.
	For each codeword $x\in \calC_b$ we add the edge $x_e= (i,\set{j,k})$ to $E$ where $x(i) = x(j) = x(k)=2$ and $i<j<k$.
	As argued previously,  edge $e\in E$ can be added by at most $2$ codewords in $\calC_b$: therefore, $|E|\geq 0.5 \times |\calC_b|$.
	Now, we will show that $G_{\calC_b}$ is $K_{s=5,t=2^{21}}$-free. 

	Suppose not, and that there exist vertices $i_1,i_2,\ldots,i_5$ in $U$ and $S_1,S_2,\ldots,S_t$ (note that each $S_i$ is a $2$-subset of coordinates) in $V$ such that $G_{\calC_b}$ induces a $K_{5,t}$ on these vertices.
	We denote the edge $(i_k, S_\ell)$ by $e_{k,\ell}$ and the codeword corresponding to it with $x_{k,\ell}$: if there are two such codewords then choose one arbitrarily.
	By the PHP there is a subset $T_1\subseteq[t]$ of size at least $t/2^4$ such that all codewords in $\set{x_{1,\ell}\mid \ell \in T}$ have the same value on the coordinates $i_2,i_3,i_4,i_5$, i.e.,  $|\set{x_{1,\ell}(i_2)\mid \ell \in T}|=1$ and so on for $i_3,i_4,i_5$.
	Again, by the PHP there is a subset $T_2\subseteq T_1$ of size at least $|T_1|/2^4=t/2^8$ such that all codewords in $\set{x_{2,\ell}\mid \ell \in T}$ have the same value on the coordinates $i_1,i_3,i_4,i_5$.
	Continuing this way for the remaining coordinates $i_3,i_4,i_5$, we obtain a subset $T_5$ of size at least $t/2^{20} = 2$:
	WLOG say $T_5=\set{S_1,S_2}$, thus, for any $c,d \in [5]$ we have $x_{c,1}(i_d) = x_{c,2}(i_d)$.
	Further, let $S = S_2\setminus S_1$, then $|S|\leq 2$. 
	By the PHP there exists $c,d\in [5]$ with $c<d$ such that $x_{c,1}(S) = x_{d,1}(S)$ where  $x(S)$ denotes the tuple $(x(i)\mid i\in S)$. 
	However, now the three codewords $x_{c,1},x_{c,2}$ and $x_{d,1}$  are a counter-example to the trifference of $\calC_b$.
	This is because for any coordinate $w\in [n]$ if $w$ is not $i_c,i_d$ or in $S_1,S_2$, then none of the three codewords have the symbol $2$ at $w$: further, if $w=i_c$, then both $x_{c,1}$ and $x_{c,2}$ have the symbol $2$ at $w$; if $w=i_d$, then, because of our definition of $T_5$ we have $x_{c,1}(w) = x_{c,2}(w)$; if $w\in S_1$, then both $x_{c,1}$ and $x_{d,1}$ have the symbol $2$ at $w$; finally if $w=S_2$, then by our assumption $x_{c,1}(S_2)=x_{d,1}(S_2)$. 
	Hence, $G_{\calC_b}$ is $K_{5,2^{21}}$-free.

	Applying \cref{thm:KST} with $u=n$ and $v=\binom{n}{2}$ and $s=5,t=2^{21}$, yields a bound on the number of edges in $E$ as $c''\times n^{3-2/5}=c''\times n^{13/5}$ for some constant $c''$.
	Hence, we obtain our desired bound on $|\calC_b|$ and $T_b(n,3)$.
\end{proof}

\begin{remark}
	We can improve the (huge) constant $2^{21}$ appearing in the above proof to some extent by following a strategy similar to the one employed in the case when $r=2$.
	However, it is easier to work with the current numbers and this doesn't seem to hurt the bounds of \cref{thm:KST} beyond a constant factor.
\end{remark}

\section{Lower Bounds on the size of \rbdtrifferentcodes{s}}
\label{sec:lower_bounds_rbdtrifferent_codes}

In this section we will prove \cref{thm:upper_bound_on_deltar}  which we restate below for convenience.

\upperboundondeltar*

We will first focus on the case of $r=1$.
\begin{lemma}[Maximum size of trifferent codes where each codeword has one $2$]
\label{lem:one_bounded_triff_codes}
    For each integer $n\geq 1$ we have $T_b(n,1)=2n$.
\end{lemma}
\begin{proof}
    As we have argued previously, in any trifferent code $\calC$, for any subset $S\subseteq [n]$ of coordinates there can be at most two codewords, say $x$ and $y$, such that $S$ is precisely the location of $2$'s for both $x$ and $y$, i.e., $S=\set{i\in [n]\mid x(i)=2}=\set{i\in [n]\mid y(i)=2}$.  Hence, $T_b(n,1)\leq 2n$.
    Next, we construct an \rbdtrifferentcodes{} ($r=1$) with block length $n$ and size $2n$. 
    Let $A_1 = \{x \in \{0,1,2\}^n \mid \# 2\text{'s in } x = 1\}$. Further, for each $i \in [n]$, let $u_i, v_i \in \{0,1,2\}^n$ be defined as 
    \[
    \begin{aligned}
    u_i(j) &= \begin{cases} 
      2 & \text{if } j = i, \\
      1 & \text{if } i < j, \\
      0 & \text{otherwise},
    \end{cases}
    &
    v_i(j) &= \begin{cases} 
      2 & \text{if } j = i, \\
      1 & \text{if } i > j, \\
      0 & \text{otherwise}.
    \end{cases}
    \end{aligned}
    \]
    Consider the code $\mathcal{C}_b \subseteq A_1$ defined as $\cup_{i \in [n]} \{u_i, v_i\}$. Clearly, $|\calC_b|=2n$. We claim the $\calC_b$ is a trifferent code. Consider any three different codewords $x,y$ and $z$: there are two cases of interest to check (other cases can be reduced to these): (a) $x=u_i,y=v_i,z=u_j\text{ or }v_j$ where $i\neq j$, in which case we have $\set{x(j),y(j),z(j)}=\set{0,1,2}$ and (b) $x=u_i,y=u_j,z=u_k\text{ or }v_k$ where $i<j<k$, in which case either $\set{x(j),y(j),z(j)}=\set{0,1,2}$ or $\set{x(i),y(i),z(i)}=\set{0,1,2}$ respectively. 
\end{proof}


Now, we turn to the general case when $r$ is a power of $3$.
\begin{lemma}[Maximum size of trifferent codes where each codeword has $3^t$ many $2$'s]
	\label{lem:upper_bound_general_rbounded}
	Let $t\geq 0$ be an integer and let $r=3^t$. Suppose that for all positive integers $n\geq r$  we have $T_b(n,r) \geq c_t \times n^{(3/2)^t}$ for some constant $c_t>0$ depending only on $t$. Then, for all positive integers $n\geq 3r$ we have $T_b(n,3r) = c_{t+1}\times n^{(3/2)^{t+1}}$ for some constant $c_{t+1}>0$ depending only on $t+1$.
\end{lemma}
\begin{proof} 
 Let $\F_q$ be a large enough finite field. 
	Let $\mathcal P = \F_q^2$ and $\mathcal L$ be the set of all the affine lines in $\F_q^2$.
	Thus,
	\begin{equation*}
		|\mathcal P| = q^2, \quad |\mathcal L| = q^2 + q.
	\end{equation*}
    Let $\mathcal S = \set{(p, \ell)\in \mathcal P\times \mathcal L \mid p\in \ell}$ and note that $|S| = q \text{ and } |\mathcal L| = q^3 + q^2$.  
	For each line $\ell \in \mathcal{L}$ choose a permutation $\sigma_\ell$ of the points of $\ell$ such that $\sigma_\ell$ has no fixed points. Further, let $f:\mathcal{S} \to \mathcal P$ be defined as $f(p,\ell) = \sigma_\ell (p)$. Notice that $f(p, \ell)$ is a point on $\ell$ but is different from $p$.
	Let $\varphi: \mathcal P \to \set{0,1,2}^n$ and $\psi: \mathcal L\to \set{0,1,2}^n$ be injective functions whose images are $3^t$-bounded trifferent codes where $n$ is the smallest integer such that $c_t\times n^{(3/2)^t} \geq q^2+q$.
	The maps $\varphi$ and $\psi$ exist by the inductive hypothesis for $t$.
	
	Define a map $\theta:\mathcal S \to \set{0,1,2}^n$ as $\theta(p, \ell) = \varphi(f(p, \ell))$.
	Finally, define $\tau:\mathcal S\to \set{0,1,2}^n\times \set{0,1,2}^n\times \set{0,1,2}^n$ as
	\begin{equation*}
		\tau(p, \ell) = (\varphi(p),\ \psi(\ell),\ \theta(p, \ell)).
	\end{equation*}
	We claim that the image of $\tau$, denoted by $\tau(\mathcal{S})$, is a $3r=3^{t+1}$-bounded trifferent code.

	Towards this end, let $e_1 = (p_1, \ell_1), e_2 = (p_2, \ell_2)$ and $e_3 = (p_3, \ell_3)$ be three pairwise distinct elements of $\mathcal S$ and consider their encodings $\tau(p_1,\ell_1),\tau(p_2,\ell_2) \text{ and } \tau(p_3,\ell_3)$. 
	If $p_1, p_2$ and $p_3$ are pairwise distinct, then $\varphi(p_1), \varphi(p_2)$ and $\varphi(p_3)$ will exhibit the trifference property as the image of $\varphi$ is a trifferent code by construction.
	Similarly, if $\ell_1, \ell_2$ and $\ell_3$ are pairwise distinct then $\psi(\ell_1), \psi(\ell_2)$  and $\psi(\ell_3)$ will exhibit the trifference property.
	So we may WLOG assume that $p_1 = p_2 = p$ (say) and $\ell_2 = \ell_3 = \ell$ (say).
	Write $p'$ in place of $p_3$ and $\ell'$ in place of $\ell_1$. See \cref{fig:two_points_two_lines}. 
    We show that $\theta(p, \ell'), \theta(p, \ell)$ and $\theta(p', \ell)$ exhibit the trifference property to finish the proof of the claim.
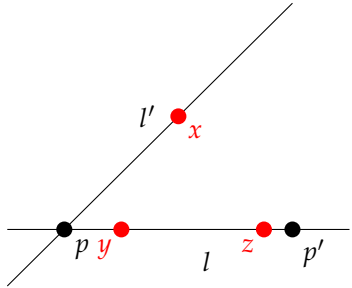
\begin{figure}[H]
\centering
\begin{tikzpicture}[scale=1.5]

  \coordinate (p) at (0,0);
  \coordinate (p') at (2,0);
  \coordinate (C) at (2,2);
  \coordinate (D) at (2.5,0);
  \coordinate (y) at (0.5,0);
  \coordinate (z) at (1.75,0);
  \coordinate (x) at (1,1);

  \draw (D) -- (p)  node[midway, below=5pt] {$l$}-- ++(-0.5,0);
  \draw (C) -- (p)  node[midway, left=5pt] {$l'$}-- ++(-0.5,-0.5);

  \fill (p) circle[radius=2pt] node[below right] {$p$};
  \fill (p') circle[radius=2pt] node[below right] {$p'$};
  \fill[red] (y) circle[radius=2pt] node[below left, text=red] {$y$}; 
  \fill[red] (z) circle[radius=2pt] node[below left, text=red] {$z$};
  \fill[red] (x) circle[radius=2pt] node[below right, text=red] {$x$};


\end{tikzpicture}
\captionof{figure}{The case when there are only two points and two lines.}
\label{fig:two_points_two_lines}
\end{figure}    

	Note that $p\neq p'$ and $\ell\neq \ell'$ for otherwise $e_1, e_2$ and $e_3$ would not be pairwise distinct.
	Let $f(p, \ell') = \sigma_\ell'(p) =  x$, $f(p, \ell) = \sigma_\ell(p) =  y$ and $f(p', \ell) = \sigma_\ell(p') =z$.
	We observe that $x, y$ and $z$ are pairwise distinct.
	Indeed, the points $y$ and $z$ lie on $\ell$ and are different from $p$: hence, $x\neq y$ and $x\neq z$ as $x$ lies on $\ell'$ and the only common point between $\ell$ and $\ell'$ is $p$. Further, $y\neq z$ because $\sigma_\ell$ is permutation of the points of $\ell$.
 
	Therefore $\varphi(x), \varphi(y)$ and $\varphi(z)$ exhibit the trifferent property.
	But
	\begin{equation*}
		\theta(e_1) = \varphi(x), \quad \theta(e_2) = \varphi(y), \quad \theta(e_3) = \varphi(z)
	\end{equation*}
	showing that $\tau(\mathcal{S})$ is a trifferent code.
	It is also clear that the number of $2$'s in each codeword of $\tau(\mathcal{S})$ is $3r$ and the block length of the code is $3n$.
	Finally, since $\tau$ is injective, the size of the code is $|\mathcal S| = q^3 + q^2$.
    Setting $n$ to be the smallest integer larger than $\left(\frac{q^2+q}{c_t}\right)^{(2/3)^t}$, and $n'=3n$, we obtain a $3r$-bounded trifferent code $\tau(\mathcal{S})$ with size at least $c_{t+1}\times n'^{(3/2)^{t+1}}$. (For smaller values of $n'>3r$ we can just adjust the constant $c_{t+1}$ to make hypothesis true.)
\end{proof}

\begin{proof}[Proof of \cref{thm:upper_bound_on_deltar}]
\cref{lem:one_bounded_triff_codes} proves the first claim.
    For the next claim: $T_b(n,3^t)\geq c_t\times n^{(3/2)^t}$ is obtained directly from \cref{lem:upper_bound_general_rbounded} with the base case of $t=0$ being \cref{lem:one_bounded_triff_codes}. Setting $r=3^t$ gives $(3/2)^t = r^\alpha$ with $\alpha = 1- \log_3(2)\approx0.369$. Hence, $\Delta_r\leq r - r^\alpha$.
\end{proof}

\bibliographystyle{alpha}
\bibliography{ref}

\end{document}